\documentclass[pre,floatfix,superscriptaddress,amsmath,amssymb,twocolumn]{revtex4}
\usepackage{framed}
\usepackage{epsfig}
\usepackage{subfigure}
\usepackage{amsmath}
\usepackage[usenames,dvipsnames]{color}
\usepackage{amssymb}
\usepackage{setspace}
\usepackage{graphicx} 
\usepackage{dcolumn}
\usepackage{bm}
\usepackage{times}
\input epsf
\usepackage{hyperref}
\hypersetup{
    bookmarks=true,         
    unicode=false,          
    pdftoolbar=true,        
    pdfmenubar=true,        
    pdffitwindow=false,     
    pdfstartview={FitH},    
    pdftitle={Efficiently inferring community structure in bipartite networks},    
    pdfauthor={Daniel Larremore, Aaron Clauset, Abigail Z. Jacobs},     
    pdfcreator={},   
    pdfproducer={}, 
    pdfkeywords={} {} {}, 
    pdfnewwindow=true,      
    colorlinks=true,       
    linkcolor=red,          
    citecolor=Green,        
    filecolor=magenta,      
    urlcolor=cyan           
}

\newcommand{\be}    {\begin{enumerate}}
\newcommand{\ee}    {\end{enumerate}}

\begin{document}

\author{Daniel B. Larremore}
\affiliation{Center for Communicable Disease Dynamics, Harvard School of Public Health, Boston, MA 02115, USA}
\affiliation{Department of Epidemiology, Harvard School of Public Health, Boston, MA 02115, USA}

\author{Aaron Clauset}
\affiliation{Department of Computer Science, University of Colorado, Boulder, CO 80309, USA}
\affiliation{Santa Fe Institute, Santa Fe, NM 87501, USA}
\affiliation{BioFrontiers Institute, University of Colorado, Boulder, CO 80303, USA}

\author{Abigail Z. Jacobs}
\affiliation{Department of Computer Science, University of Colorado, Boulder, CO 80309, USA}

\begin{abstract}
	Bipartite networks are a common type of network data in which there are two types of vertices, and only vertices of different types can be connected.
	While bipartite networks exhibit community structure like their unipartite counterparts, existing approaches to bipartite community detection have drawbacks, including implicit parameter choices, loss of information through one-mode projections, and lack of interpretability.
	Here we solve the community detection problem for bipartite networks by formulating a bipartite stochastic block model, which explicitly includes vertex type information and may be trivially extended to $k$-partite networks. 
	This bipartite stochastic block model yields a projection-free and statistically principled method for community detection that makes clear assumptions and parameter choices and yields interpretable results.
	We demonstrate this model's ability to efficiently and accurately find community structure in synthetic bipartite networks with known structure and in real-world bipartite networks with unknown structure, and we characterize its performance in practical contexts.
	
\end{abstract}

\title{Efficiently inferring community structure in bipartite networks}

\maketitle

\section{Introduction}

The defining feature of a bipartite network is that there are two types of vertices, $a$ and $b$, and only vertices of different types may be connected; there are no edges connecting vertices of the same type. For example, if type $a$ vertices represent flowers and type $b$ vertices represent pollinating insects, two vertices $i$ and $j$ are connected if flower $i$ is pollinated by insect $j$; two flowers will never be connected, nor will two insects. Bipartite networks appear specialized but are remarkably common. Examples include networks of plants and pollinators \cite{bascompte2006}, as well as documents and words \cite{dhillon2001,lancichinetti2014}, genes and genetic sequences \cite{larremore2013}, actors and movies \cite{peixoto2013,peixoto2014,baconnumber}, social network users and mobile access locations \cite{ye2011}, and scientific papers and their authors \cite{newman2001,newman2002,newman2003,erdosnumber}. 

As with unipartite networks, a common task is to find groups or communities of vertices that connect to the rest of the network in similar ways. Finding this underlying group structure has many uses, including dividing a heterogeneous network into more homogeneous subgraphs for subsequent analysis or modeling. However, communities in bipartite networks do not fit the commonly-used definitions. Such definitions are usually motivated by assortative community structure in social networks~\cite{newman2003}, where vertices in the same community are more likely to be connected than vertices of different communities. In a bipartite network, however, two vertices of the same type can never be connected, and thus assortativity-based definitions of communities are ill-suited. In this paper, we present a bipartite formulation of the popular stochastic block model, which provides a statistically principled solution to the community detection problem for bipartite networks and defines a community as a group of vertices with similar connectivity patterns to other groups.

Common approaches to community detection in bipartite networks include applying standard community-detection algorithms to a one-mode projection \cite{zhou2007}. In a one-mode projection, two type $a$ vertices are connected if they share a common type $b$ neighbor. By eliminating all type $b$ vertices, this procedure effectively reduces the dimensionality of the network by discarding information. Often, projections are created implicitly, without first constructing the bipartite network. For instance, in a scientific coauthorship network, a pair of authors are connected if they ever wrote a paper together \cite{newman2001,newman2002,newman2003}, which is a one-mode projection of the larger bipartite network of all papers and authors. Measures like the Erd\H{o}s number \cite{erdosnumber} or Bacon number \cite{baconnumber} are, in fact, counting path lengths on projections of bipartite networks.
 
Using projections creates both practical and principled issues. Projections are necessarily composed only of overlapping cliques, which are extremely low probability under most community detection null models, including Girvan-Newman modularity $Q$~\cite{girvan2002}, and tend to inflate measures such as assortativity and the clustering coefficient. Moreover, reducing the effective dimensionality of the data almost always requires a loss of information; not only can structurally different bipartite networks exhibit identical one-mode projections \cite{zhou2007}, but even the projection of a highly structured bipartite network can appear unstructured, which we demonstrate in our results. 

To avoid these issues, two bipartite extensions of Girvan-Newman modularity \cite{girvan2002} have been proposed. Broadly speaking, one approach formulates a null model for vertices connected to each other in the projection \cite{guimera2007}, while the other formulates a null model for vertices connected to each other in the bipartite network \cite{barber2007}. Both express implicit modeling restrictions and assumptions in their outputs: maximizing the modularity of Guimera {\it et al.}\ partitions one type of vertex at a time so that each type's partition is independent of the other \cite{guimera2007}, while maximizing Barber's modularity yields mixed-type groups (i.e., groups that consist of vertices of both types) \cite{barber2007}.  Other methods find pure-type groups while using the full bipartite network, and are sometimes called co-clustering or co-partitioning methods \cite{dhillon2001}.

Stochastic block models (SBMs) are elegant probabilistic models of group structure in networks~\cite{holland1983,wang1987,nowicki2001,karrer2011,shen2011,decelle2012,peixoto2013,peixoto2014} that have been used to identify community structure in biological networks~\cite{allesina2009,larremore2013}, product recommendation systems~\cite{guimera2012}, and  directed social cooperation networks \cite{rovira-asenjo2013}. SBMs are often capable of community detection in bipartite networks \cite{karrer2011,shen2011,peixoto2013,peixoto2014}, and some SBM-based schemes have been developed for the specific case of bipartite networks with multiple non-overlapping edge types \cite{guimera2012,rovira-asenjo2013}. 

Generally, however, SBMs are \textit{generative models} for networks with block or community structure, meaning one can partition the vertices into $K$ groups, specify the connectivity parameters among groups, and then generate network data. In this way, the SBM defines a parametric probability distribution over all networks. When given a network, community detection becomes a form of inference, in which we aim to find the parameters that best explain observed network data, which is equivalent to finding configurations that minimize the system's free energy. Relative to many other community detection techniques, stochastic block models have the advantage of explicitly stating the underlying assumptions, which improves the interpretability of the results.

In fact, we may specify parameters for a SBM that will produce bipartite networks, and for this reason, community detection in bipartite networks is possible by directly applying the SBM to bipartite data. We may also apply the SBM to one-mode projections of bipartite networks. However, we will show later that, even though the SBM is flexible enough to accommodate both of these cases, the bipartite formulation of the SBM exhibits both improved speed and improved quality of community detection.

In the following sections we formulate the bipartite stochastic block model (biSBM) and describe an algorithm that searches for a maximum likelihood partition of a network into communities. We first show that the biSBM can correctly extract a planted network partition from a noisy background, particularly in a case where the one-mode projection is uninformative. We then apply the biSBM to several empirical networks, showing that the biSBM outperforms its non-bipartite SBM counterpart.

\section{The Bipartite Stochastic Block model}\label{formulation}

Our approach to the bipartite stochastic block model, hereafter biSBM, builds on recent work of Karrer and Newman \cite{karrer2011}, who described a simple SBM that generates networks with a fixed expected degree sequence. This degree-corrected SBM is substantially more effective at finding a correct partition when vertex degrees are heterogeneous, as in many real-world networks. We first introduce the simple case, and then extend it to include degree correction. 

We begin by dividing the $N_a$ vertices of type $a$ into $K_a$ groups and the $N_b$ vertices of type $b$ into $K_b$ groups. In this way, each group or community contains vertices of a single type. We use the $N \times N$ adjacency matrix $A$ rather than the $N_{a} \times N_{b}$ bipartite adjacency matrix $B$, which are related as
\[ A = \left ( \begin{array}{cc}
0 & B \\
B^{\top} & 0 \end{array} \right ) \enspace . \]
Similarly, we express the matrix of group interrelationships $\omega$ as a $K \times K$ matrix (where $K = K_a + K_b$), instead of a $K_a \times K_b$ matrix, as is sometimes chosen. We will set to zero any entries of $A$ and $\omega$ that would connect vertices of the same type, thereby enforcing bipartite structure. This notation is more easily extended to $k$-partite or more complicated networks, is less cumbersome, and is consistent with previous work on the SBM \cite{karrer2011}. 

Let vertex $i$ be of type $t_{i}$ and belong to group $g_{i}$. Let $T_{r}$ be the type of group $r$, imposing the constraint
\begin{equation}
	t_{i} = T_{g_{i}}\enspace, 
	\label{eq-bipartiteconstraint}
\end{equation}
which indicates that vertex types and group types must match and ensures that groups will be pure-type. With this common set of definitions, we develop the biSBM without and with degree correction.

\subsection{biSBM without degree correction}

The block structure of the biSBM network is defined by a $K \times K$ matrix $\omega$. Let $\omega_{rs}$ be the expected value of the adjacency matrix entry $A_{ij}$ for vertices $i$ and $j$ belonging to groups $r$ and $s$ respectively. Let the number of actual edges between $i$ and $j$ be drawn from a Poisson distribution with the corresponding mean. Although most real-world networks do not have multi-edges, we allow them here because the Poisson distribution makes calculations easier, and because for sparse networks in which $\omega_{rs}$ is small, multi-edges are highly unlikely and corrections to the simpler Bernoulli probabilities become vanishingly small. Enforcing the bipartite constraint of Eq.~\eqref{eq-bipartiteconstraint} produces a restriction on $\omega$
\begin{equation}
	\omega_{rs} = 0 \qquad \text{ when } T_r = T_s \enspace .
	\label{eq-omegazeros}
\end{equation}

This equation restricts the model to bipartite networks only, in both generation and inference. When presented with a bipartite network, the lack of edges between vertices of the same type is not informative to the biSBM; it is taken as a given. The SBM, on the other hand, makes no such assumption. The lack of edges between subsets of vertices is informative to the SBM, and so it must discover bipartite structure from the data and weigh a bipartite partition against non-bipartite alternatives. We discuss this point in more detail in Sec.~\ref{discussion}. 

Given parameters $g$, $T$, and $\omega$, we can write down the probability of generating a network $G$ with adjacency matrix $A$
\begin{equation}
	P(G\,|\ g,\omega,T) = \prod_{\substack{i<j\\ t_{i} \neq t_{j}}} \frac{\left(\omega_{g_{i}g_{j}}\right)^{A_{ij}}}{A_{ij}!} \exp{\!\left(-\omega_{g_{i}g_{j}} \right)} \enspace.
	\label{eq-bsbm1}
\end{equation}
By using the symmetry of $A$ and $\omega$, this can be rewritten as
\begin{equation}
	P(G\,|\ g, \omega, T) = \!\!\prod_{\substack{i<j\\ t_{i} \neq t_{j}}} \!\frac{1}{ A_{ij} !} \times \!\!\prod_{\substack{r\ s \\ T_{r} \neq T_{s}}} \!\omega_{rs}^{m_{rs}/2} \exp{\!\left(-\frac{1}{2} n_{r} n_{s} \omega_{rs}\right)} \enspace ,
	\label{eq-bipprob}
\end{equation}
where $n_r$ is the number of vertices in group $r$ and $m_{rs}$ is the number of edges between groups $r$ and $s$, defined using the Kronecker $\delta$ function as
\begin{equation}
	m_{rs} = \sum_{i\ j} A_{ij} \ \delta_{g_{i},r}\ \delta_{g_{j},s} \enspace .
	\label{eq-mrs}
\end{equation}
Given a bipartite network $G$ with adjacency matrix $A$ and vertex types $t$ \cite{note-types}, we seek the parameters that maximize Eq.~\eqref{eq-bipprob}. In practice, it is easier to maximize its logarithm, since this changes only the value of the maximum but not its location in parameter space. Neglecting constants, taking the log yields
\begin{equation}
	\ln{P(G\,|\ g, \omega)} = \sum_{\substack{r\ s \\ T_{r} \neq T_{s}}} m_{rs} \ln{\omega_{rs}} - n_{r}n_{s}\omega_{rs} \enspace .
	\label{eq-logbipprob}
\end{equation}
Following Ref. \cite{karrer2011}, we maximize this sum first with respect to $\omega$ and then with respect to $g$. Taking a derivative of Eq.~\eqref{eq-logbipprob} with respect to $\omega_{rs}$ and setting it equal to zero yields
\begin{equation}
	\hat{\omega}_{rs} = \frac{m_{rs}}{n_{r}n_{s}} \enspace .
\end{equation}
A variable with caret denotes a maximum likelihood parameter estimate, while one without denotes a model parameter. Substituting this expression into Eq.~\eqref{eq-logbipprob} yields
\begin{equation}
	\ln{P(G\,|\ g, \hat{\omega}, T)} = \sum_{\substack{r\ s \\ T_{r} \neq T_{s}}} m_{rs} \ln{ \frac{m_{rs}}{n_{r} n_{s}}} - m_{rs} \enspace, 
\end{equation}
where the latter term sums to twice the number of edges in the network, regardless of the partition. We therefore drop it, yielding
\begin{equation}
	\mathcal{L}(G\,|\ g) = \sum_{\substack{r\ s \\ T_{r} \neq T_{s}}} m_{rs} \ln{\frac{m_{rs}}{n_{r}n_{s}}}\enspace ,
	\label{eq-bsbm}
\end{equation}
which we now maximize over all group assignments $g$, subject to the constraint of Eq.~\eqref{eq-bipartiteconstraint} which requires that any partition $g$ must divide vertices into pure-type communities. 

\subsection{Degree-corrected biSBM}

Both the motivation for and derivation of the degree-corrected biSBM parallel those of the degree-corrected SBM: real-world networks tend to have broad degree distributions in addition to community structure, but the uncorrected biSBM finds edge bundles between communities with Poisson degree distributions, which in practice tends to sort vertices by degree. The degree-corrected model explicitly models the observed degree sequence before finding community structure, allowing it to be applied to empirical networks with broad degree distributions. 

As before, we consider a network of $N$ vertices, indexed by $i$, each with type $t_i$, divided into $K_a$ type $a$ groups and $K_b$ type $b$ groups, with $g_i$ denoting the group to which vertex $i$ belongs. Let $\theta_i$ control the expected degree of vertex $i$, and let $\omega_{rs}$ again be a $K \times K$ symmetric matrix of parameters to control the number of edges between groups $r$ and $s$. Following \cite{coja-oghlan2010}, we let the numbers of edges between vertices $i$ and $j$ follows a Poisson distribution with mean $\theta_{i} \theta_{j} \omega_{g_{i}} \omega_{g_{j}}$. To enforce the bipartite structure of the network, Eqs.~\eqref{eq-bipartiteconstraint} and \eqref{eq-omegazeros} must hold, and the probability of observing a network $G$ with adjacency matrix $A$ is 
\begin{equation}
	P(G\,|\, g,\theta, \omega, T) = \!\!\prod_{\substack{i<j\\ t_{i} \neq t_{j}}} \frac{\left(\theta_{i} \theta_{j}\omega_{g_{i}g_{j}}\right)^{A_{ij}}}{A_{ij}!} \exp{\!\left(-\theta_{i} \theta_{j}\omega_{g_{i}g_{j}} \right)}\enspace .
	\label{eq-dcsbm1}
\end{equation}
The parameters $\theta$ are arbitrary to within a multiplicative constant that can be absorbed into $\omega$, so we choose the normalization
\begin{equation}
	 \sum_{i} \theta_{i} \delta_{g_{i},r} = 1\enspace ,
	 \label{eq-thetaconstraint}
 \end{equation}
which means that $\theta_{i}$ is the probability that an edge connected to the community to which vertex $i$ belongs lands on $i$ itself. This constraint allows Eq.~\eqref{eq-dcsbm1} to be rewritten as
\begin{equation}
	P(G\,|\, g, \theta, \omega, T) = \frac{\prod_{i} \theta_{i}^{k_i}}{\prod_{\substack{i<j\\ t_{i} \neq t_{j}}} A_{ij} !} \times \!\!\! \prod_{\substack{r\ s \\ T_{r} \neq T_{s}}} \!\!\omega_{rs}^{m_{rs}/2} \exp{\!\left(-\frac{1}{2} \omega_{rs} \right)} ,
	\label{eq-bipprobdc}
\end{equation}
where $k_i$ is the observed degree of vertex $i$ and $m_{rs}$ is the number of edges between groups $r$ and $s$, as before [Eq.~\eqref{eq-mrs}]. We again seek to maximize this probability by maximizing its logarithm. After dropping constants and multiplying by 2, we have
\begin{equation}
	\ln P(G\,|\, g,\theta,\omega) = 2 \sum_{i} k_{i} \ln{\theta_{i}} + \!\!\! \sum_{\substack{r\ s \\ T_{r} \neq T_{s}}} \!\! m_{rs} \ln{\omega_{rs}} - \omega_{rs} \enspace .
	\label{eq-dcmaximizeme}
\end{equation} 
Taking partial derivatives with respect to $\omega_{rs}$ and setting them equal to zero gives the maximum likelihood parameters
\begin{equation}
	\hat{\omega}_{rs} = m_{rs}\enspace .
\end{equation}
The maximum likelihood $\hat{\theta}_i$ can be found via the constrained maximization of Eq.~\eqref{eq-dcmaximizeme} subject to Eq.~\eqref{eq-thetaconstraint} using Lagrange multipliers, yielding
\begin{equation}
	\hat{\theta}_i = \frac{k_{i}}{\kappa_{g_i}}\enspace ,
\end{equation}
where $\kappa_r$ is the sum of the degrees in group $r$, $\kappa_r = \sum_{s} m_{rs}$. The maximum likelihood parameter estimates preserve not only the expected numbers of edges between groups, but also the expected degree sequence of the network \cite{karrer2011}. They may be substituted into Eq.~\eqref{eq-dcmaximizeme}, and after manipulation and dropping constant terms, we have
\begin{equation}
	\mathcal{L}(G\,|\,g) = \sum_{\substack{r\ s \\ T_{r} \neq T_{s}}} m_{rs} \ln \frac{m_{rs}}{\kappa_{r} \kappa_{s}}\enspace , 
	\label{eq-dcbsbm}
\end{equation} 
which we maximize over all partitions $g$. 

As in the case of non-bipartite networks, the differences between the uncorrected and corrected log-likelihood functions, Eqs.~\eqref{eq-bsbm} and \eqref{eq-dcbsbm} respectively, appear to be a simple substitution of $n_r$ for $\kappa_r$, but their effect on optimal partitions can be drastic when degrees are heterogeneous, which we will demonstrate in Sec.~\ref{results}. Both formulations of the model will find $K$ pure-type groups, $K_a$ within the vertices of type $a$ and $K_b$ within the vertices of type $b$.

\subsection{A biSBM algorithm}\label{algorithm}

To maximize Eq.~\eqref{eq-bsbm} or \eqref{eq-dcbsbm} over all partitions $g$, we present an algorithm adapted from Karrer and Newman's algorithm~\cite{karrer2011}, which is a variation on the classic Kernighan-Lin algorithm \cite{kernighan1970}. Our algorithm takes as inputs the adjacency matrix $A$ and the vertex types $t_i$, and then assigns vertices of type $t_i=a$ uniformly at random to $K_a$ groups, indexed $\{1,\dots,K_a\}$, and vertices of type $t_i=b$ uniformly at random to $K_b$ groups, indexed $\{K_a+1,\dots,K_a+K_b\}$. This means $T_r=a$ for the first $K_a$ groups, and $T_r=b$ for the remaining $K_b$. 

The algorithm searches the likelihood surface by proposing to move a vertex from one group $r$ to another group $s$, provided their types match $T_r = T_s$. After proposing all such moves, across all vertices and eligible groups, it selects the move that will most increase the likelihood function. If no improvement is possible, the algorithm chooses the move that least decreases the likelihood function, because forcing the vertices to move helps escape local optima \cite{note-k1relaxation}. We allow each vertex to move only once, and when all vertices have moved, the states through which the system has passed are evaluated and the state with the highest objective score is used as a starting point for the next search iteration. When a full iteration passes with no improvement in objective score, the algorithm exits. 

Finally, as is usual with stochastic optimization techniques, the algorithm should be run many times and the highest score from among these independent replicates selected. This algorithm may be used equally well for the degree-corrected or uncorrected models. 
	
\section{Comparison of the biSBM and SBM}\label{discussion}

Before demonstrating that the bipartite stochastic block model correctly extracts community structure from bipartite network data, we first examine the relationship between the biSBM and the SBM. Most SBM community detection methods can be naturally applied to bipartite networks \cite{karrer2011, decelle2012, peixoto2013,peixoto2014}, so it may not be clear why a specialized bipartite model is necessary. In this section, we characterize the relationship between the biSBM and the SBM both theoretically and in application, showing that the models are related but do not perform equivalently. In particular, the SBM often overfits bipartite data by mixing vertices of different types within communities and it is nearly always substantially less efficient.
 
\subsection{Relationship to the non-bipartite stochastic block model}\label{relationshiptosbm}

The derivation of the biSBM requires that there be no connections between any vertices of the same type. We expressed this in Eqs.~\eqref{eq-bipartiteconstraint} and \eqref{eq-omegazeros}, and formulated the biSBM equations accordingly. We now show that if these two constraints are applied {\it a posteriori} to the SBM and degree-corrected SBM, the resulting equations will be numerically equal to the biSBM;  any network that is generated by the (degree-corrected) biSBM can be generated with equal probability by the properly constrained (degree-corrected) SBM. Indeed, it is well known that stochastic block models are capable of producing bipartite networks, in addition to general multipartite networks \cite{karrer2011, decelle2012}, and networks with more complicated rules about which types of vertices may be connected to which other types, so this equivalence of generative models comes as no surprise. 

The biSBM and degree-corrected biSBM likelihood functions are numerically equivalent to their non-bipartite counterparts, provided that (i) the partition $g$ does not mix vertices of different types in the same group, and (ii) there are no edges between vertices of the same type. To see this, we reproduce the probability of generating a graph $G$ with adjacency matrix $A$ using the SBM from Ref.~\cite{karrer2011}
\begin{align}
	P(G\,| \omega, g) = & \prod_{i<j} \frac{(\omega_{g_{i} g_{j}})^{A_{ij}}}{A_{ij}!} \exp{(-\omega_{g_{i}g_{j}})} \nonumber \\
	& \times \prod_{i} \frac{(\frac{1}{2}\omega_{g_{i} g_{i}})^{A_{ii}/2}}{(A_{ii}/2)!} \exp{\left(-\frac{1}{2}\omega_{g_{i}g_{i}}\right)} \enspace .
	\label{eq-probsbm}
\end{align}
If there are no edges between groups of the same type, then $\omega_{g_{i}g_{i}}=0$, so every term in the second product is equal to one and may be disregarded. Moreover, $\omega_{g_{i}g_{j}}=0$ when $i$ and $j$ are of the same type, so all terms of the remaining product equal one, except those for which $t_i \neq t_j$, which reduces numerically to Eq.~\eqref{eq-bsbm1}. However, these equations, while numerically equivalent, are not identical due to their meanings and behaviors.

The ability of the SBM to generate the same ensemble of bipartite networks as the biSBM does not imply that they will find identical partitions when presented with real data. There are two reasons for this, one principled and one algorithmic. The key to both is understanding the way that each model makes use of the data presented to it. Equation~\eqref{eq-omegazeros} means that the lack of edges between vertices of the same type is uninformative to the biSBM because it is taken as a given. On the other hand, to the SBM, the lack of edges between vertices of the same type {\em is} informative to the model, which uses such information for inference. 

In other words, the likelihood function for both models is determined by the density of observed edges between the communities of the partition, and this function is maximized whenever the density parameter is close to $1$ or $0$. Thus, the SBM prefers to find either very assortative or very disassortative groups, or some mixture thereof, while the biSBM can find only disassortative groups, by definition. Thus, when applied to bipartite data, the SBM must learn that all groups are in fact disassortative, while the biSBM does not.

The structure of the objective function produces a strong incentive for the SBM to find disassortative structure in bipartite networks, but this incentive is not sufficient to always find pure-type partitions in bipartite data. As we show below, for many simple bipartite networks, a mixed-type partition in which vertices of different types are placed in the same group yield a higher likelihood than pure-type (bipartite) partitions for the SBM. (After all, the biSBM and SBM are nested models, and thus the SBM can always find a parametrization at least as good as that of the biSBM.) 

\begin{figure}
	\centering
	\includegraphics[width=1.0\linewidth]{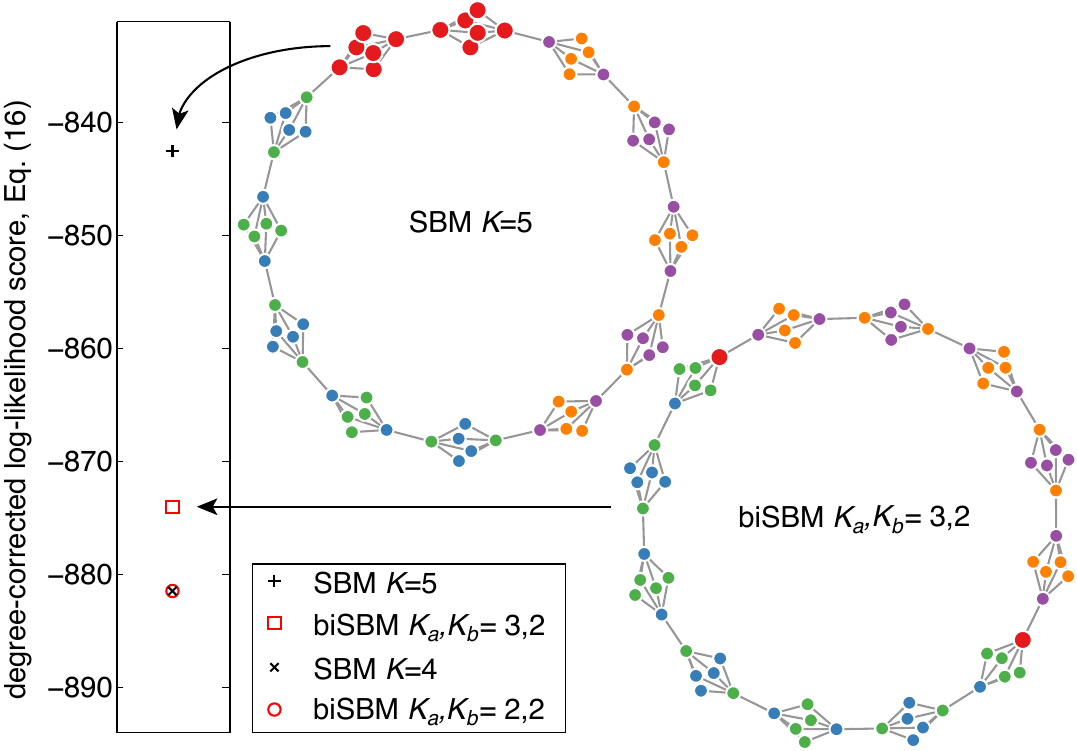}
	\caption{(Color online) The global-maximum partitions of the biSBM and SBM are not necessarily the same. When $K$ is even, the SBM and biSBM find identical partitions, but when $K$ is odd, the SBM finds a higher likelihood partition by creating a mixed-type community. Log-likehoods are plotted, and partitions are displayed as colors, with the mixed-type partition vertices (red) enlarged.}
	\label{fig-letograph_k} 
\end{figure}

To illustrate this point, consider a simple network consisting of a ring of small ``clumps,'' each of which is a perfectly bipartite structure (Fig.~\ref{fig-letograph_k}). Whenever $K$ is odd, the SBM will overfit by finding a partition that mixes vertex types but which also has a higher objective score than the best bipartite partition under the biSBM. Whenever $K$ is even, the SBM and biSBM find identical partitions. While this illustrates the point that the maximum likelihood partition under the SBM may be better than that under the biSBM, the SBM finds a bipartite partition for as much of the network as possible until it is forced to break symmetry by the $K=5$ specification. These results hold for both degree-corrected and uncorrected models.

\subsection{Performance relative to SBM}
Since we have just established that it is possible for the SBM to find higher likelihood partitions than the biSBM without providing $t$, the vertex type information, one might prefer community detection using the SBM because it requires less information and is more flexible. However, we now demonstrate that for even moderate $N$ or $K$, the biSBM finds better solutions, faster. This occurs because the biSBM simultaneously solves two smaller problems, one for each vertex type, and because the ruggedness of the likelihood surface presents the SBM with many more local optima in which it can become lodged.

 \begin{figure}
	\centering
	\includegraphics[width=1.0\linewidth]{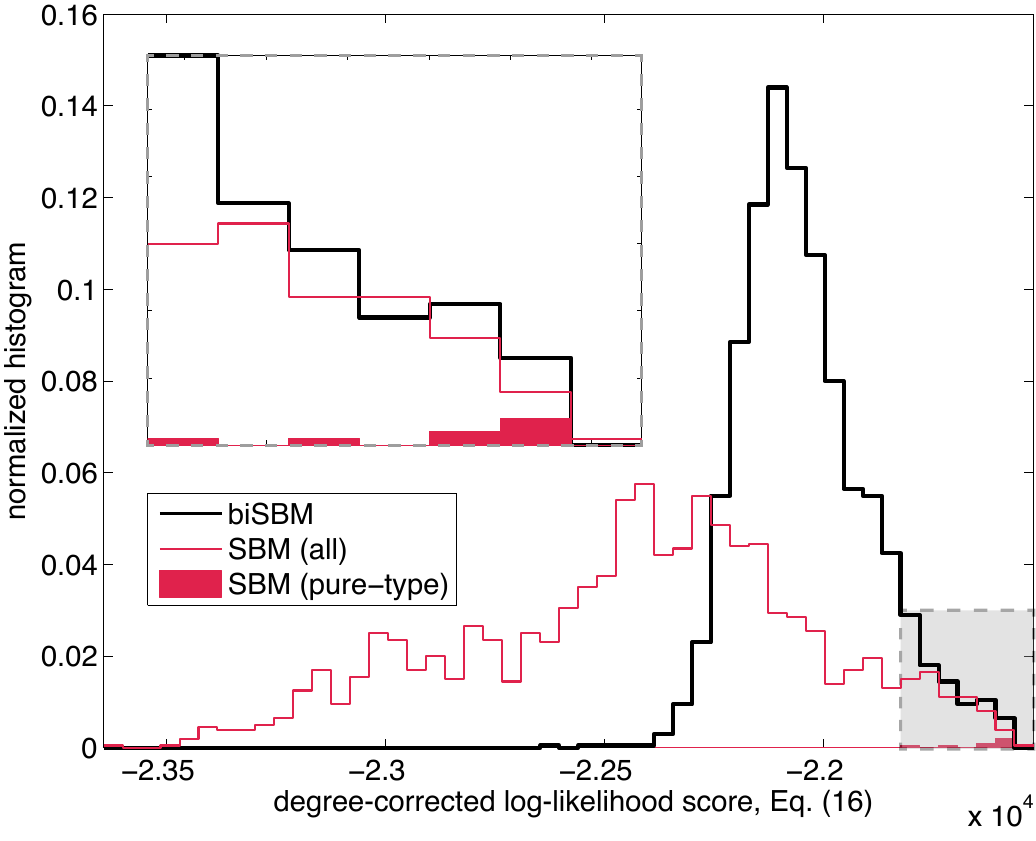}
	\caption{(Color online) The biSBM outperforms the SBM in speed, log-likelihood score, and the ability to find partitions that do not mix vertices of different types (pure-type). The inset magnifies the shaded region of the main plot which includes all eight pure-type partitions (of 2000 total replicates) found by the SBM. Times to convergence for each replicate were $5.33$ and $1.64$ seconds for the SBM and biSBM respectively. Tests were conducted using the malaria dataset (see text) and $K_a,K_b=3,3$ and $K=6$.}
	\label{fig-speed} 
\end{figure}

We compare our biSBM algorithm with the SBM algorithm on which it was based, provided by Karrer and Newman~\cite{karrer2011}. They describe the change in likelihood $\Delta \mathcal{L}$ of moving a vertex $i$ from community $r$ to community $s$, and explain that this quantity can be evaluated for the degree-corrected model in time $\mathcal{O}(K+\langle k \rangle)$ per vertex on average, where $\langle k \rangle$ is the mean degree. Thus, finding the community $s$ that is the very best move for vertex $i$ takes $\mathcal{O}[K(K+\langle k \rangle)]$ time. Overall, the time complexity of the SBM is
\begin{equation}
	\mathcal{O}\big[NK(K+\langle k \rangle)\big]\enspace .
	\label{eq-speed-sbm}
\end{equation}
The biSBM algorithm separates $N$ searches over $K$ communities into $N_a$ searches over $K_a$ communities and $N_b$ searches over $K_b$ communities. The time complexity of each biSBM iteration is therefore roughly
\begin{equation}
	\mathcal{O}\big[ N_a K_a(K_a+\langle k \rangle)\big] + 	\mathcal{O}\big[N_b K_b(K_b+\langle k \rangle)\big] \enspace .
	\label{eq-speed-bsbm}
\end{equation}
By using $K = K_a + K_b$, and $N=N_a+N_b$, and the fact that $(x+y)^{2} \geq x^{2} + y^{2}$ for $x,y \geq 0$, one can show that the biSBM is always faster than the SBM, in large part because the biSBM's search space is predivided by vertex type into two smaller problems.

Applying the degree-corrected SBM and biSBM algorithms to a dataset from the genes of the malaria parasite (described in detail in Sec.~\ref{malaria}), we plot the final log-likelihood scores for each of $2000$ iterations as histograms for each method in Fig.~\ref{fig-speed}. The results show that the biSBM tends to find better partitions than the SBM in each iteration, and the SBM rarely finds pure-type partitions (eight of 2000 replicates). Moreover, we find that the biSBM converges 3.25 times faster than the SBM, which took $5.33$ seconds per replicate.

The difference in times arises from Eqs.~\eqref{eq-speed-sbm} and \eqref{eq-speed-bsbm}, while the difference in outcomes is due to the high-dimensional ruggedness in the SBM's likelihood function. On this function, most random initializations lie within the basin of attraction of a local optimum corresponding to a mixed-type partition with a lower log-likelihood. In contrast, by eliminating all mixed-type partitions, the biSBM restricts the search and guides the optimization to generally higher-quality solutions. We note that the popular modularity score $Q$ for assortative community detection exhibits a qualitatively similar rugged structure, with many local optima and a complex distribution of basins of attraction~\cite{good2010}.

As a final test, we examined the stability of biSBM partitions under the SBM algorithm to determine whether the SBM's additional flexibility in parameter space would allow for an improved partition. In all cases considered, when initialized at a partition found by the biSBM, this partition was also a local optimum for the SBM. This behavior suggests that the biSBM's smaller parameter space provides a significant speed advantage over the SBM, without any tradeoff in partition quality, i.e., good optima of the biSBM are also good optima of the SBM.

\section{Results}\label{results}

\begin{figure*}
	\centering
	\includegraphics[width=0.33\linewidth]{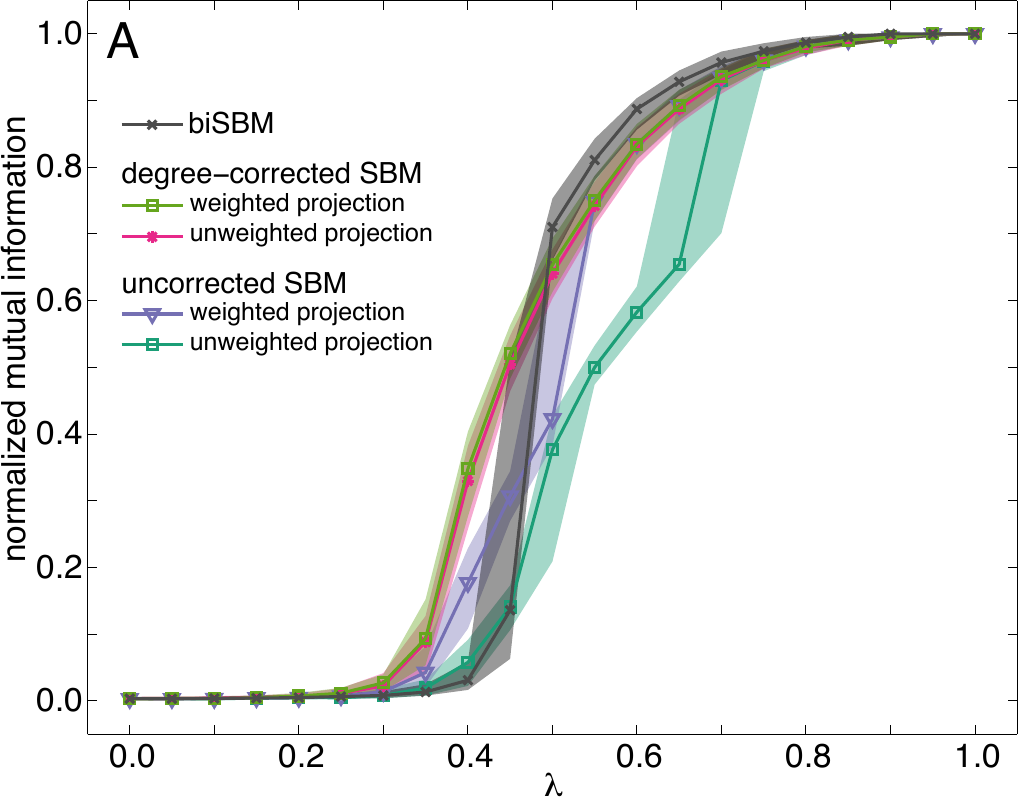}
	\includegraphics[width=0.33\linewidth]{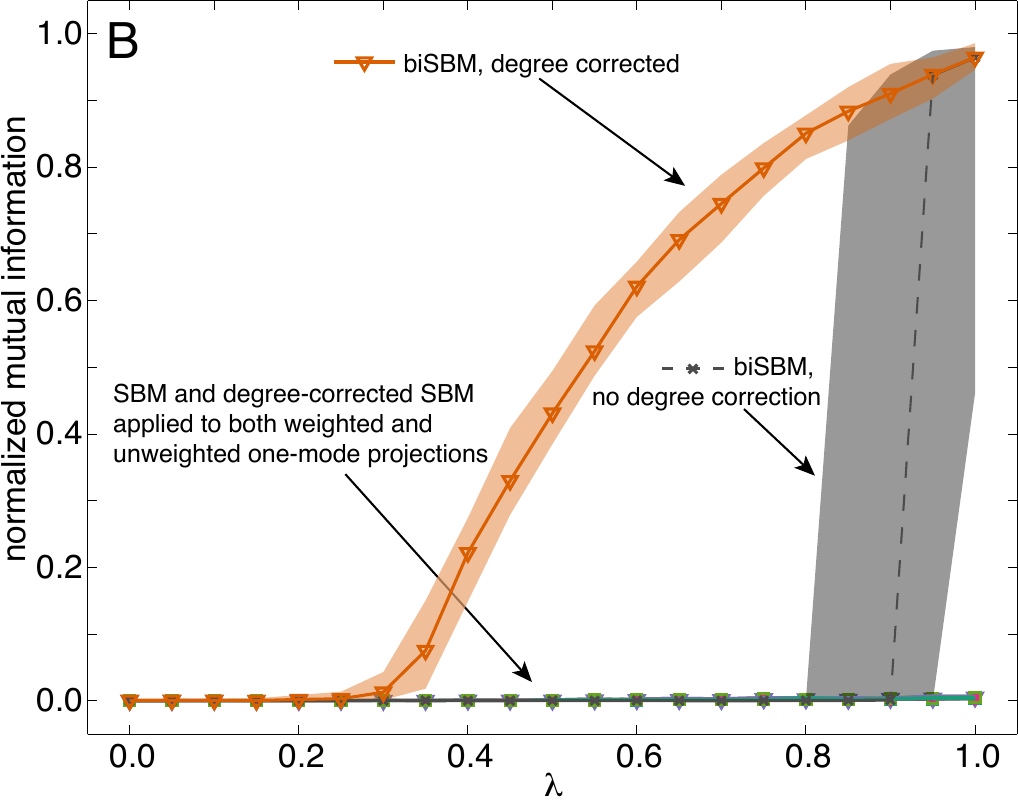}
	\includegraphics[width=0.33\linewidth]{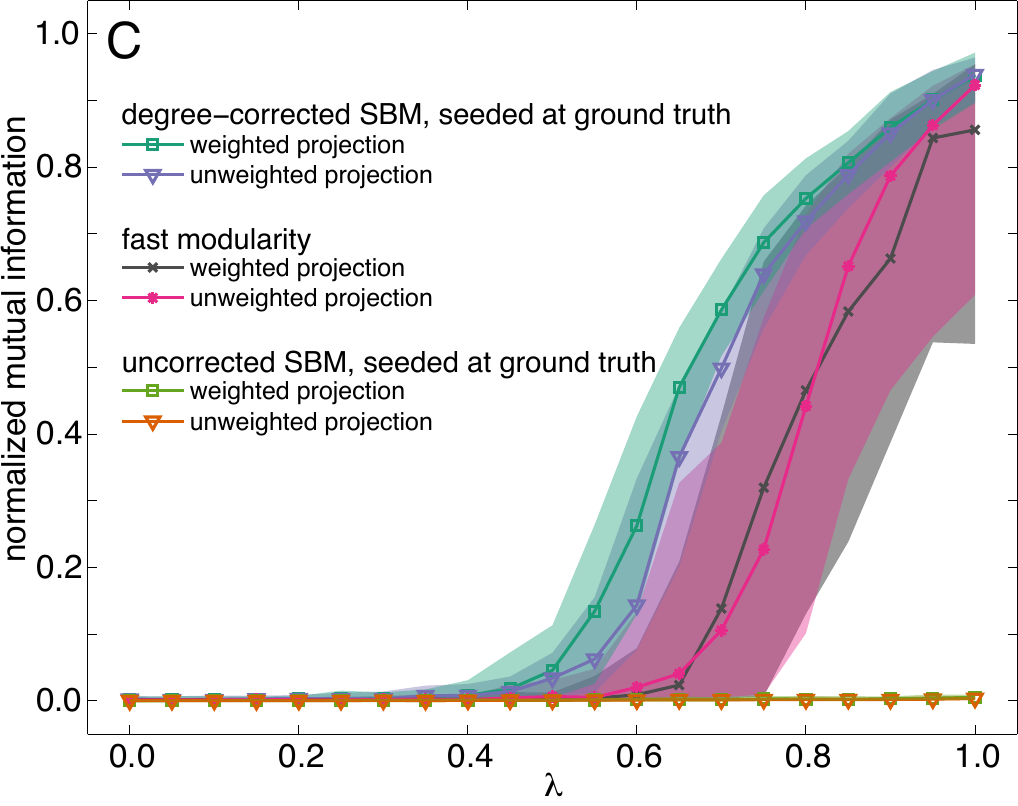}
	\caption{(Color online) As the level of noise is decreased ($\lambda\to1$), $I_\text{norm}$ between inferred and correct partitions varies by method. Each point shows the median of 100 replicates, and shaded regions show 10\%--90\% quantiles. (A) In the easy case, all methods are able to find the correct partition. The degree-corrected SBM applied to projections performs slightly better for small $\lambda$ and the biSBM performs slightly better for moderate and large $\lambda$. (B) In the difficult case, only the degree-corrected biSBM is able to reliably find the correct partition; SBM methods applied to projections failed. (C) For the same projections as panel B, fast modularity maximization is moderately accurate but inconsistent. When initialized at the correct partition, the degree-corrected SBM remains nearby in parameter space for large $\lambda$ but the uncorrected SBM does not.}
	\label{fig-synthetic2}
\end{figure*}

In this section, we show that the biSBM can recover the correct partition in synthetic networks with known planted structure and then apply the biSBM to study three empirical networks. For the synthetic networks, we consider two forms, an easy and a hard case, which illustrate the biSBM's performance under different general conditions and against alternative techniques.
Of the empirical data sets, the first is the Southern Women network \cite{davis1941}, which consists of 18 women who attended 14 social events. This network is commonly used as a benchmark for bipartite network community detection algorithms, much like the Zachary karate club for unipartite community detection algorithms. Past work in this direction, while agreeing broadly on a partition of the women \cite{freeman2003,guimera2007}, says little about a partition of events (except \cite{barber2007}). The biSBM provides both. The second is the malaria network, which consists of genetic sequences from the malaria parasite {\it Plasmodium falciparum} \cite{rask2007,larremore2013}.  Its vertices correspond to 297 genes and their 806 shared amino acid substrings, and projections of similar networks have been previously analyzed \cite{bull2008,larremore2013}. The third network is a subset of the Internet Movie Database (IMDb) network of actors and movies, consisting of $53,158$ actors and the $39,768$ movies in which they appear. 

\subsection{Synthetic Networks}

We examine the ability of the algorithm to extract planted structure $\omega^{\text{planted}}$ that has been obscured by various levels of uniformly random noise. Empirically observed networks are often noisy, with missing or spurious edges, and a good community detection algorithm must be able to extract structure despite such a noisy background. 

We describe two forms of synthetic networks, each of which illustrates a different aspect of community detection in bipartite networks. The first form is easy, because it consists of four equally sized, unambiguous, and non-overlapping components, each made up of one type $a$ and one type $b$ community. In this case, community structure is obvious in both the bipartite network and its one-mode projection.  The second form is difficult because, in addition to $K_a \neq K_b$, its degrees and community sizes are heterogeneous. Moreover, its one-mode projection is ambiguous and difficult to resolve, even in the absence of noise. Here, only the degree-corrected biSBM correctly finds the planted community structure.  These two forms are not exhaustive but rather illustrate the practical behavior of the biSBM.

To vary the amount of noise, we specify $g$ and $\omega^{\text{planted}}$ but create networks using $g$ and $\omega = \lambda \omega^{\text{planted}} + (1-\lambda) \omega^{\text{random}}$, letting the mixing parameter $\lambda$ take values between $0$ (all noise) and $1$ (all planted structure). The construction of $\omega^{\text{random}}$ depends on whether we use the degree-corrected or uncorrected model. In the uncorrected model, we preserve the expected number of edges in the network, but remove all structure, and thus $\omega^{\text{random}}_{rs} = n_{r}n_{s}/2m$, where $m$ is the total number of edges in the network. In the degree-corrected model, we preserve both the expected number of edges in the network and the expected degrees of the vertices $\theta$, and thus $\omega^{\text{random}}_{rs} = \kappa_{r}\kappa_{s}/2m$.

To further illustrate the point that one-mode projections induce practical issues for community detection in bipartite networks, we also compare partitions of one-mode projections of our synthetic networks with the performance of the biSBM. There are two types of such projections. An unweighted projection of a bipartite network onto its type $a$ vertices is obtained by letting two type $a$ vertices $i$ and $j$ be connected if they share any type $b$ neighbor $k$. Each edge of a weighted projection has weight equal to the number of such shared neighbors. Given an adjacency matrix $A$, the weighted projection matrix $P$ is given by 
\begin{equation}
	P = A^2 ,
\end{equation}
where the diagonal blocks of size $N_a \times N_a$ and $N_b \times N_b$ correspond to the projections onto types $a$ and $b$ vertices, respectively. The matrix $P$ is equivalent to a ``two-step'' adjacency matrix, with each entry weighted by the number of length-2 paths between each pair of vertices.

In our experiments, performance is evaluated by specifying parameters to the biSBM, drawing network instances from that ensemble, and then testing a method's ability to recover the correct partition of type $b$ vertices. This allows a direct comparison of the biSBM (which partitions all vertices) and the SBM (which partitions only type $b$ vertices). Accuracy is measured by the normalized mutual information between the inferred and correct partitions~\cite{danon2005}. We treat each partition as a random variable $X$. Since the only information we have about $X$ is what we observe, let $\text{Pr}(X=r) = N_{r}/N$, the fraction of vertices observed in group $r$. Similarly, let the joint distribution of two partitions $X$ and $Y$ be defined as $\text{Pr}(X=r,Y=s) = N_{rs}/N$, the fraction of vertices that we observe in group $r$ of the first partition and group $s$ of the second partition. Then, the normalized mutual information of the partitions is $I_{\text{norm}}(X,Y) = 2 I(X, Y) / \left [ H(X) + H(Y) \right]$, where $H(X)$ is the Shannon entropy of $X$, and $I(X,Y)$ is the mutual information. As the name implies, $I_{\text{norm}}$ takes on values between $0$ and $1$, with $I_{\text{norm}}(X,Y)=1$ if and only if $X=Y$, and $I_{\text{norm}}=0$ when $X$ and $Y$ are uncorrelated. Intuitively, $I_{\text{norm}}(X,Y)$ measures the degree to which knowledge of one partition allows us to predict the other partition.

\subsubsection{An easy case}\label{easy}

In this easy case, we define the mixing matrix to have easily identifiable community structure
\begin{equation}
	\omega^{\text{planted}} = 
	\begin{pmatrix}
\cdot & \cdot & \cdot & \cdot &		\alpha & 0 & 0 & 0 \\
\cdot & \cdot & \cdot & \cdot &		0 & \beta & 0 & 0 \\
\cdot & \cdot & \cdot & \cdot &		0 & 0 & \gamma & 0 \\
\cdot & \cdot & \cdot & \cdot &		0 & 0 & 0 & \delta \\ 
\alpha & 0 & 0 & 0 &		\cdot & \cdot & \cdot & \cdot  \\
0 & \beta & 0 & 0 &		\cdot & \cdot & \cdot & \cdot  \\
0 & 0 & \gamma & 0 &		\cdot & \cdot & \cdot & \cdot  \\
0 & 0 & 0 & \delta &		\cdot & \cdot & \cdot & \cdot  \\
	\end{pmatrix}\enspace ,
\end{equation}
where the variables $\alpha,\beta,\gamma,\delta$ are positive constants. This produces a network with four components, each consisting of a pair of communities. We let $N=1000$ for each type and divide these vertices evenly across the four components. Finally, we do not specify vertex degrees $\theta$, and thus create networks using $\omega^{\text{random}}$ for the uncorrected SBM. 

For this test, we compare the performance of the biSBM on bipartite data to the performance of the SBM on both weighted and unweighted one-mode projections, which simulates the common practice of converting bipartite data into a form amenable to standard unipartite detection methods. Figure~\ref{fig-synthetic2}A shows the normalized mutual information between the inferred partitions of type $a$ vertices and the correct partition of type $a$ vertices. The biSBM always extracts the correct communities when $\lambda=1$, with performance falling off sharply as the network approaches the detectability limit~\cite{decelle2012} where no algorithm can recover the planted structure. In this case, because the structure is unambiguous, projection methods also work well. 

\subsubsection{A difficult case}\label{difficult}

\begin{figure}
	\centering
	\includegraphics[width=0.7\linewidth]{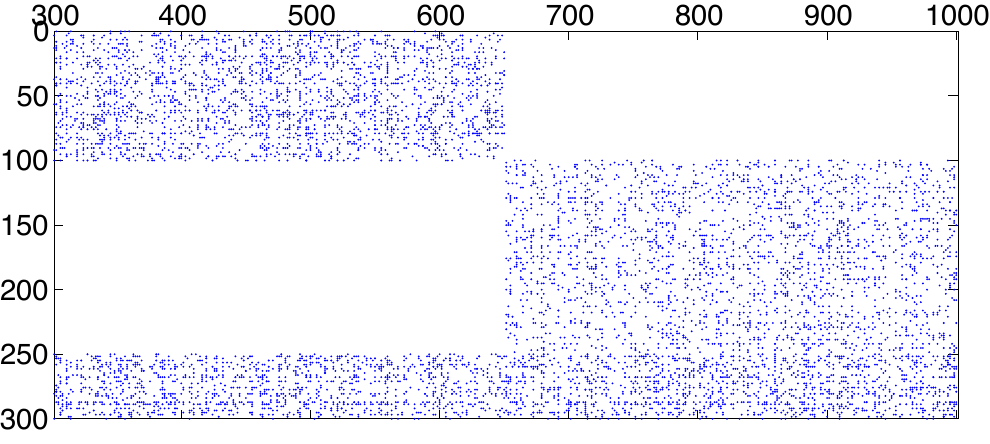}
	\includegraphics[width=0.7\linewidth]{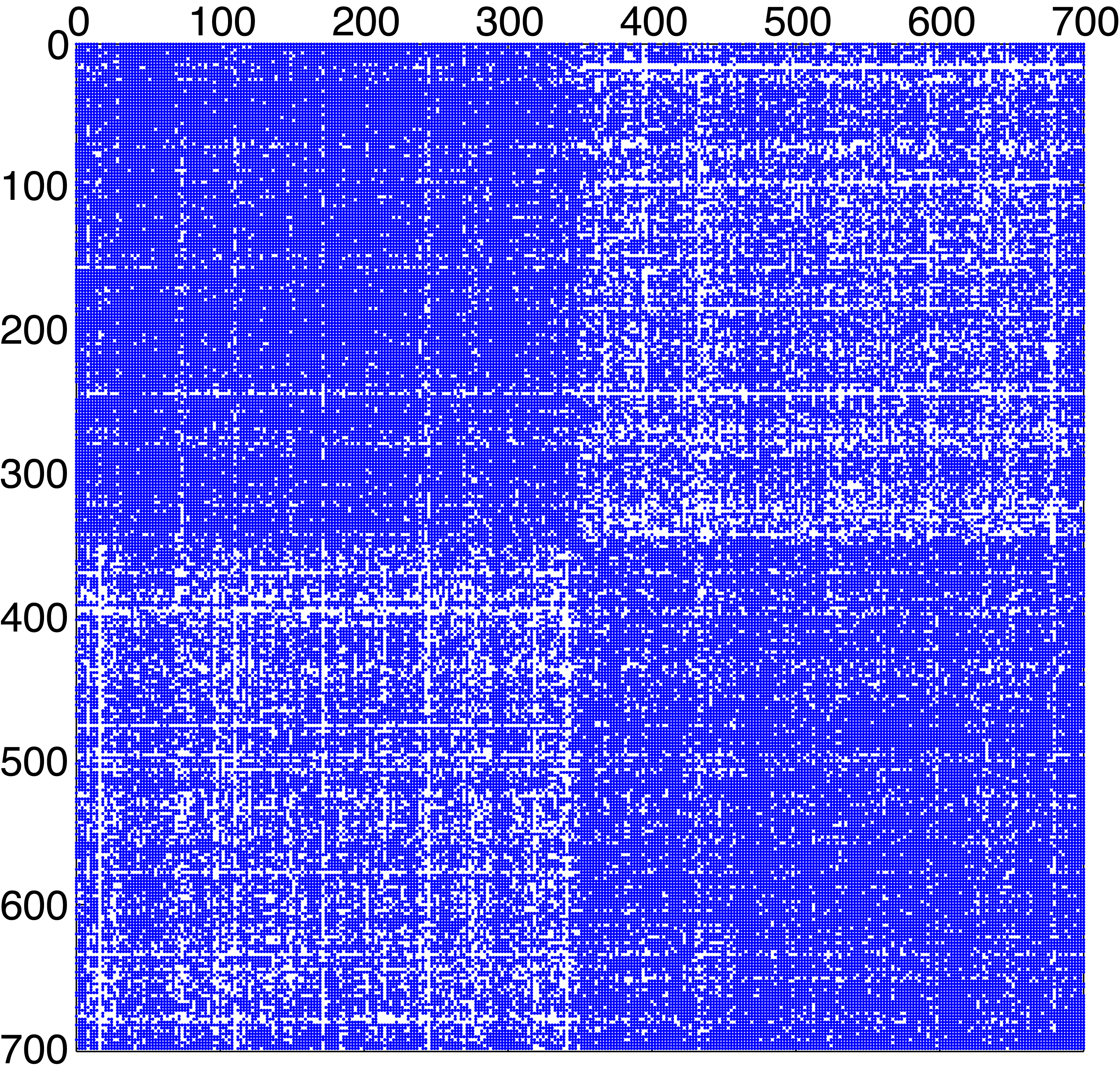}
	\caption{(top) The bipartite adjacency matrix $B$ for the planted structure Eq.~\eqref{eq-ambig}. (bottom) The $b$-mode projection exhibits visible community structure when correctly sorted, which is undetectable by the SBM (see Fig.~\ref{fig-synthetic2}B).}
	\label{fig-synthetic2_adj}
\end{figure}

In this difficult case, we define the mixing matrix to have less easily identifiable community structure by creating partially overlapping communities, $K_a \neq K_b$, and a broad degree distribution. Moreover, we illustrate this in a network whose one-mode projection is relatively uninformative about its community structure
\begin{equation}
	\omega^{\text{planted}} =  
	\begin{pmatrix}
\cdot & \cdot & \cdot &		\epsilon & 0 \\
\cdot & \cdot & \cdot &		0 & \epsilon \\
\cdot & \cdot & \cdot &		\gamma &  \gamma \\
\epsilon & 0 & \gamma & \cdot & \cdot \\
0 & \epsilon & \gamma & \cdot & \cdot
	\end{pmatrix} \enspace .
	\label{eq-ambig} 
\end{equation}
In this construction, the third type $a$ community connects equally with both type $b$ communities. When the network is projected onto its type $b$ vertices, this equality masks much of the structure created by the other, non-overlapping type $a$ communities, making the projection difficult to partition, even when $\gamma \sim \epsilon$. We make this test even more difficult for the biSBM by choosing different sizes for the communities \cite{danon2006,rosvall2007}, with $300$ type $a$ vertices, divided $\{100,150,50\}$, and $700$ type $b$ vertices divided evenly $\{350, 350\}$. Finally, we impose heterogeneous degrees by giving half the vertices in each community twice the preferred degree $\theta$ of the others \cite{noteOnDegrees}. As such, we use $\omega^{\text{planted}}$ corresponding to a random network with fixed degree sequence. To clearly illustrate the planted structure of the bipartite adjacency matrix, we plot one such matrix for $\lambda = 1$ in Fig.~\ref{fig-synthetic2_adj}, and show its type $b$ projection. 

Figure~\ref{fig-synthetic2}B shows the normalized mutual information between the inferred partitions of type $b$ vertices alone. The degree-corrected biSBM exhibits the classic detectability phase transition~\cite{decelle2012}, with a critical point at $\lambda\approx0.33$. In contrast, the uncorrected biSBM finds the planted structure only for $\lambda\approx1$, but as shown by the 10\% and 90\% quantiles (shaded regions), its partitions are either extremely accurate or extremely inaccurate. 

When using either weighted or unweighted projections, the SBM (with or without degree correction) is unable to find any community structure. Ordering the adjacency matrix by the planted partition, however, shows clear community structure (Fig.~\ref{fig-synthetic2_adj}), which the SBM algorithm is unable to find. Initializing the SBM algorithm with the correct partition does lead to better performance (Fig.~\ref{fig-synthetic2}C) for the degree-corrected SBM, which remains near the correct partition when $\lambda \approx 1$, while the uncorrected SBM fails completely. This indicates that the correct partition of the projection is not a local optimum under the uncorrected SBM.

Corroborating a result for bipartite modularity maximization \cite{guimera2007}, the weighted projection outperforms the unweighted projection in this experiment. Figure~\ref{fig-synthetic2}C also shows that fast modularity maximization \cite{clauset2004} is able to partially extract structure from the projection, but with high variability. This suggests that the projection's communities for $\lambda > 0.5$ are not below the detectability limit \cite{decelle2012}, but that they are nevertheless very difficult to find, highlighting a case in which applications of community detection to projections are outperformed by the biSBM.

While this bipartite network was designed to produce a relatively uninformative projection, it represents a common type of bipartite network in which some vertices have a very high degree. Such networks arise in document classification, when words are connected to the documents in which they are found, because some words, such as {\it up}, {\it again}, and {\it which}, appear frequently, and without any correlation to topics. Bipartite co-clustering methods have been shown to succeed even when such ``stop words'' are included \cite{dhillon2001}, but projection-based methods require removal of  these words because they effectively mask the true structure in uncorrelated noise \cite{lancichinetti2014}. Bipartite methods will therefore be particularly useful in contexts where the list of stop words is not known {\it a priori}.

\begin{figure}
	\centering
	\includegraphics[width=0.8\linewidth]{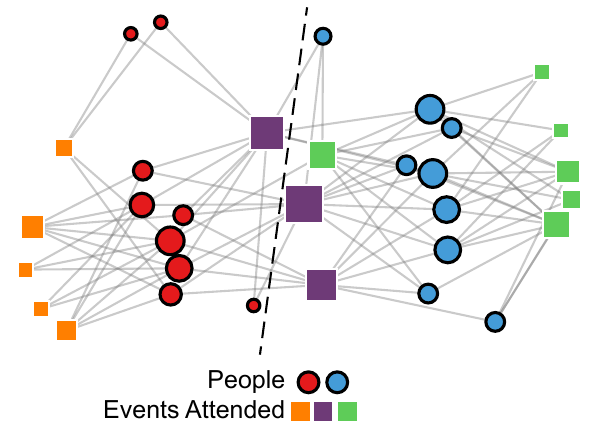}
	\caption{(Color online) The bipartite SBM correctly classifies the women (circles) of the Southern Women data set \cite{davis1941}. Vertex area is proportional to degree, and colors label the partition, with black outlines corresponding to women and white outlines corresponding to events (squares). Degree correction does not have an effect on the maximum likelihood partition for this network. The dashed line corresponds to the two-community partition found in Ref.~\cite{guimera2007}, which separately partitioned women and events.}
	\label{fig-southernWomen}
\end{figure}

\subsection{Empirical Networks}

\begin{figure*}[ht]
	\centering
	\includegraphics[width=0.48\linewidth]{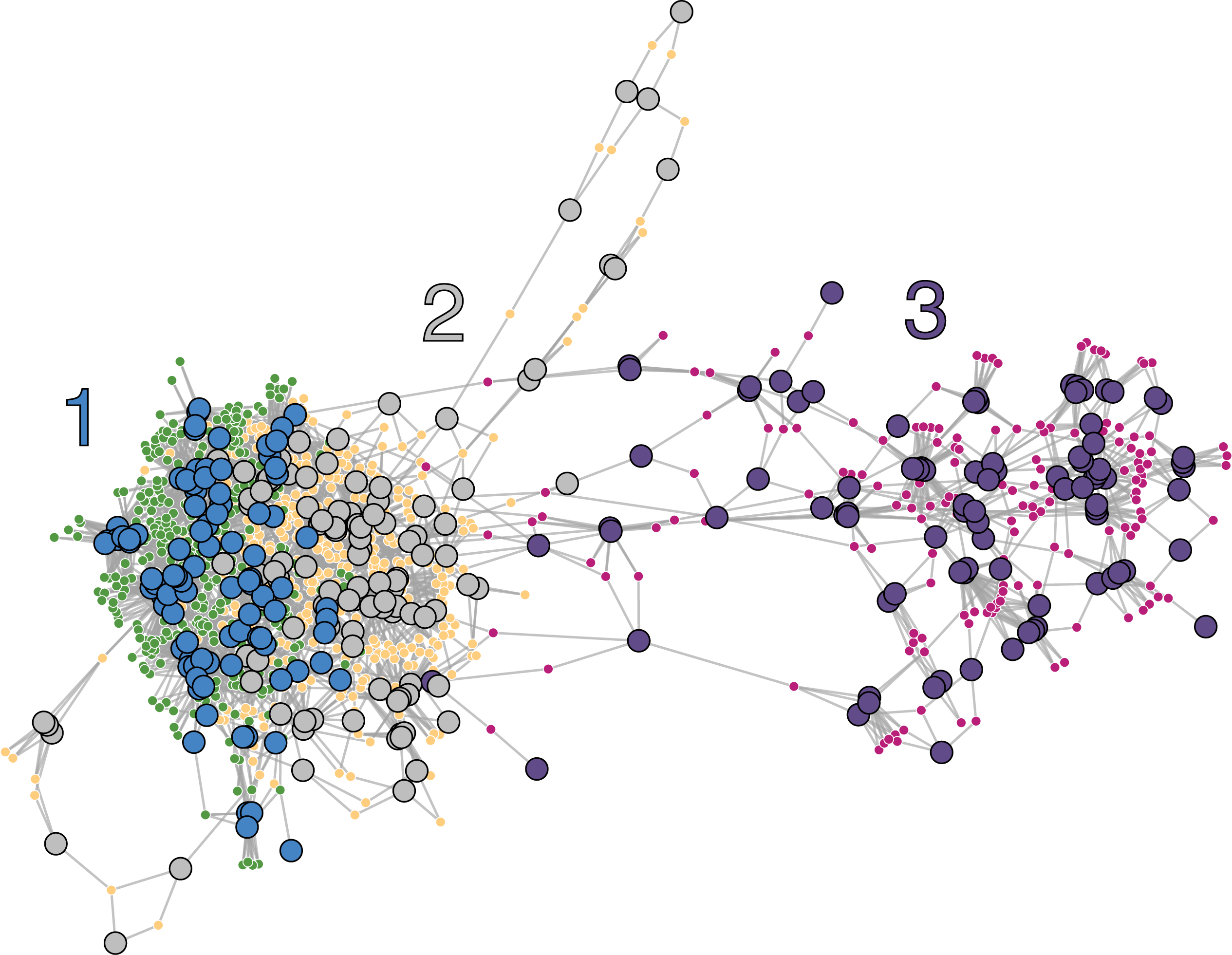}
	\includegraphics[width=0.48\linewidth]{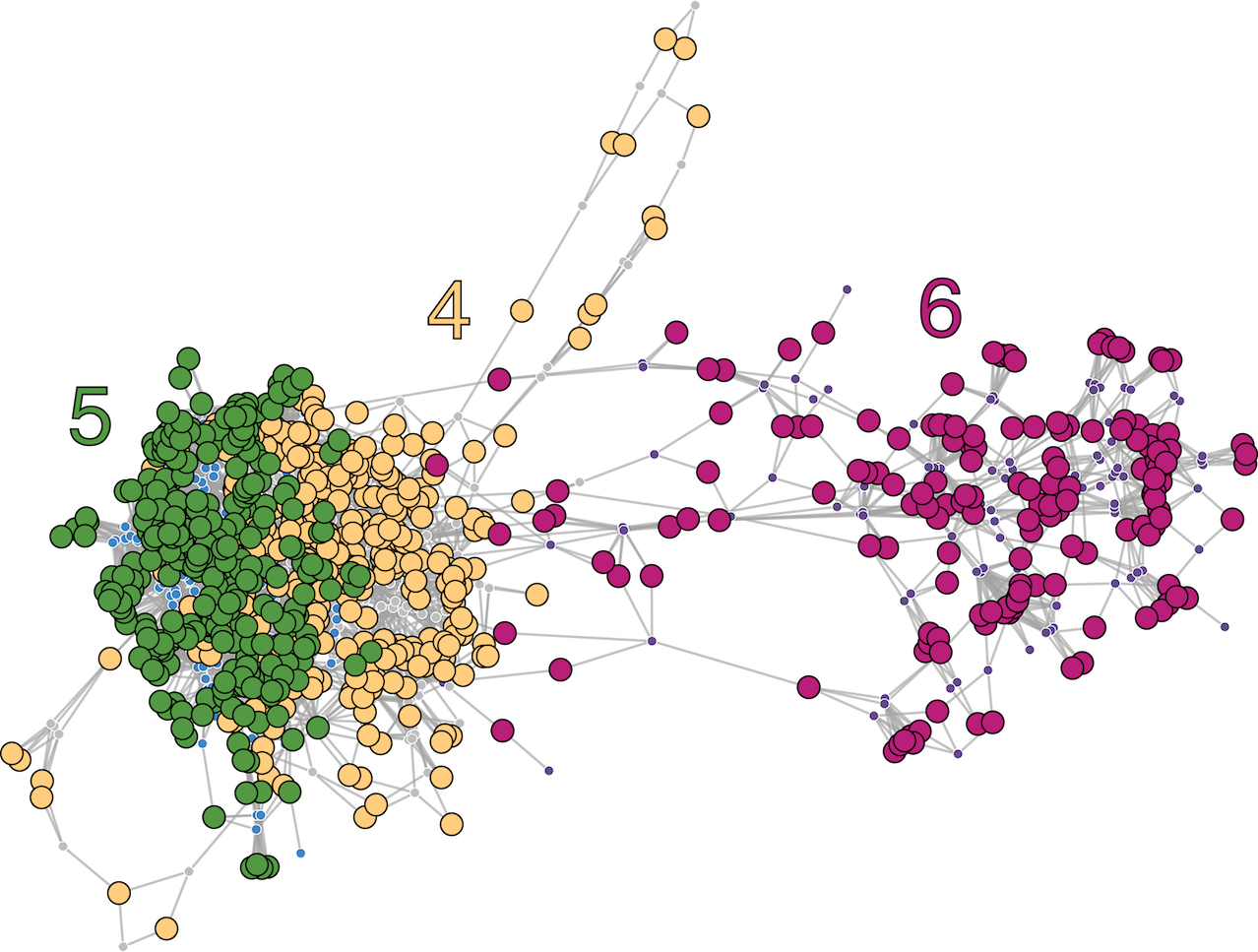}
	\caption{(Color online) The force-directed layout of the malaria bipartite network is shown twice, with gene-vertices enlarged (left) and with substring-vertices enlarged (right). Numbers and colors indicate the partition found by the degree-corrected biSBM for $K_{a} = 3$, $K_{b}=3$. The paired communities on the right side of the figures (3 and 6) are almost non-overlapping with the others, which are partially overlapping. The corresponding bipartite adjacency matrix is shown in Fig.~\ref{fig-malaria_adj}.}
	\label{fig-malaria}
\end{figure*}

\subsubsection{The Southern Women Dataset}

Our first empirical network is the Southern Women dataset, a common benchmark for bipartite community detection algorithms \cite{barber2007,guimera2007}. It reflects attendance at 14 social events by 18 women in Natchez, Mississippi, USA in the 1930s, and the data were collected by ethnographers to examine the roles of race and class in dictating social interactions \cite{davis1941,freeman2003}.

The biSBM and degree-corrected biSBM identified the same partition, shown in Fig.~\ref{fig-southernWomen}. The partition of women perfectly matched the literature consensus \cite{freeman2003} and that of Guimera {\it et al.}~\cite{guimera2007}. The partition of events found by Guimera {\it et al.}, shown as the dashed line in Fig.~\ref{fig-southernWomen}, split events into two groups, largely matching the three group partition that we show. Barber's modularity was maximized with four mixed-type communities \cite{barber2007}, although the consensus partition noted above has only a slightly worse modularity. Our partition is listed explicitly in Appendix \ref{appendix-southernWomen}.

In this example, the biSBM performs well and is able to find the literature consensus partition of the women while simultaneously partitioning events. However, this dataset serves as a minimal benchmark: although 21 different methods were reviewed in Ref.~\cite{freeman2003}, a majority produced identical partitions, with many of the others differing by a single vertex label. Therefore, in the next section, we present the biSBM with a more challenging empirical network. 

\subsubsection{Malaria Dataset}\label{malaria}

Our second empirical network comes from the malaria parasite {\it P. falciparum}. The parasite evades the human immune system via a protein camouflage, which is encoded in {\it var} genes \cite{note-var}. In order to create novel camouflages, {\it var} genes frequently recombine, which amounts to the constrained splicing and shuffling of genetic substrings, giving rise to community structures naturally \cite{larremore2013,bull2008}. Vertex types correspond to genes and their constituent substrings, and each substring connects to every gene in which it is present. The network, consisting of $297$ genes and $806$ substrings, is somewhat like a set of documents and words, but with partially overlapping words, and covers a subset of the known {\it var} genes. Degree distributions for both types of vertices are broad which makes it an exemplar for the degree-corrected biSBM.

\begin{figure}[t]
	\centering
	\includegraphics[width=1.0\linewidth]{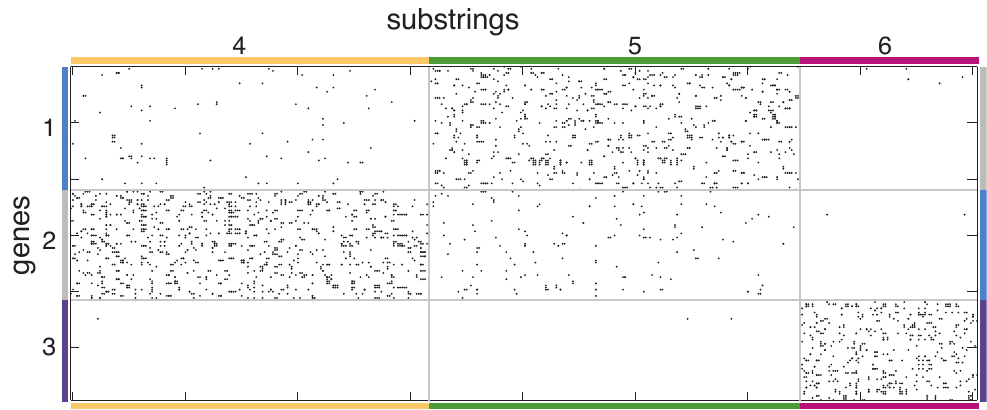}
	\caption{(Color online) The bipartite adjacency matrix $B$ of the malaria network, sorted by the degree-corrected biSBM partition, $K_{a}=3, K_{b}=3$. Numbers and colors on the matrix border correspond to those in Fig.~\ref{fig-malaria}.}
	\label{fig-malaria_adj}
\end{figure}

Sample partitions using $K_{a}=3$, $K_{b} = 3$ are shown in a force-directed layout in Fig.~\ref{fig-malaria}. The degree-corrected biSBM recovers communities of different sizes, as shown in the plotted adjacency matrix, Fig.~\ref{fig-malaria_adj}. One group of genes corresponds nearly exclusively to one group of substrings, while the other two groups of genes and substrings are partially overlapping. Community sizes and degrees vary by community but are easily accommodated by the degree-corrected biSBM. A superset of these data were analyzed previously \cite{larremore2013}, finding a similar partition of the genes, but no partition of the substrings. See Appendix~\ref{codeanddata} for the data and partition. 

To illustrate the difference between degree-corrected and uncorrected models, we also applied the uncorrected biSBM to the malaria dataset, and found that connected vertices tended to group by degree, corroborating analogous findings for the non-bipartite SBM \cite{karrer2011}. Moreover, the maximum likelihood partition, which we plot in Fig.~\ref{fig-malariaNODC}, does not correspond well to biological classifications of the genes \cite{larremore2013}. As with the synthetic networks in the previous subsection, when networks have broad or heterogeneous degree distributions, the degree corrected model is able to find the correct partition while the uncorrected model is not. 

\subsubsection{IMDb Dataset}\label{imdb}

Our third empirical network comes from the Internet Movie Database (IMDb), from which we built a bipartite network of actors and the movies in which they acted. Data were downloaded directly from IMDb \cite{imdb} and parsed into a network in which an edge exists between an actor and a movie if the actor was in the movie in any role. We removed all serial television shows included in the database, restricted the network to movies released between 1995 and 2000, and then removed any actor or movie with degree equal to one, as in other studies \cite{peixoto2013,peixoto2014}. From this, we extracted the largest connected component, resulting in a single-component network of $53,158$ actors and $39,768$ movies. Degree distributions for both vertex types were broad, with mean degrees of 7.6 and 5.7, and maximum degrees of $120$ and $552$, for movies and actors, respectively.  

\begin{figure}[t]
	\centering
	\includegraphics[width=1.0\linewidth]{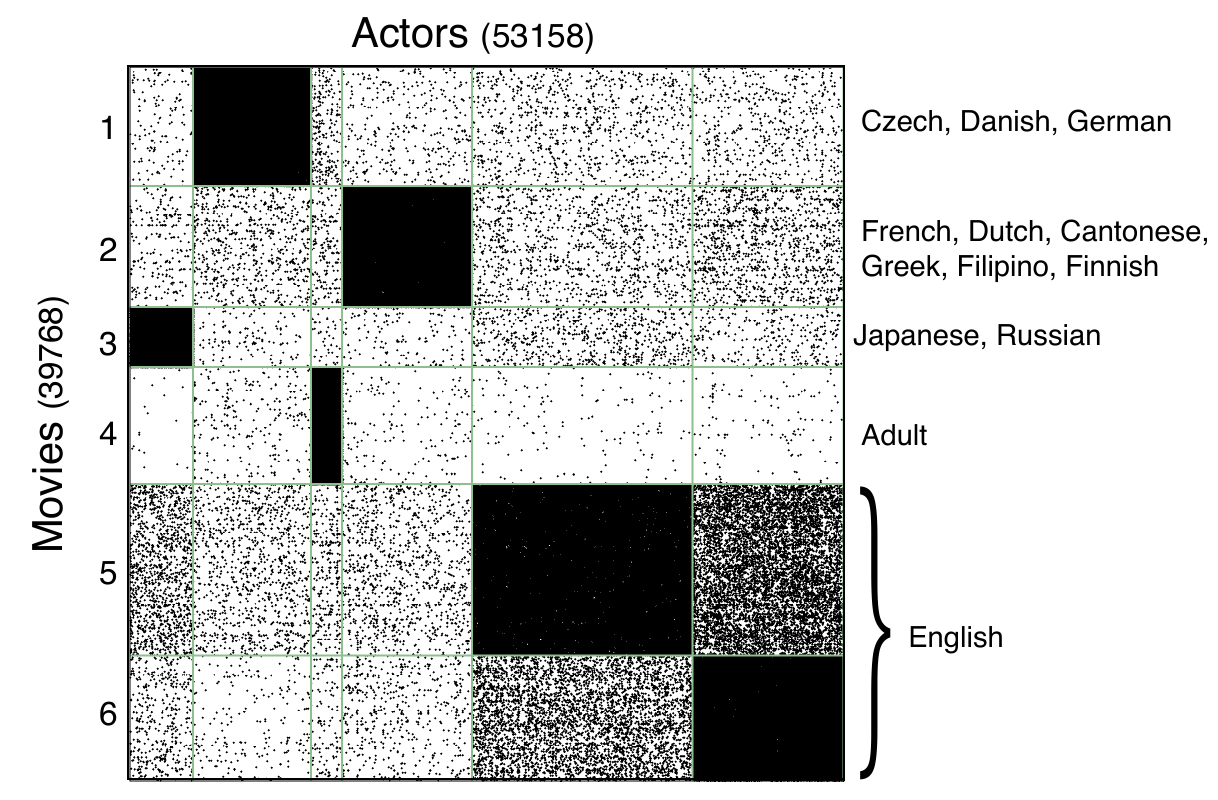}
	\caption{(Color online) The bipartite adjacency matrix $B$ of the IMDb network \cite{imdb}, sorted by the degree-corrected biSBM partition with $K_{a}=6, K_{b}=6$. Language labels indicate that over 90\% of movies in the indicated language are in that group. Group $4$ is best characterized by the Adult genre, and features a much larger number of movies per actor in the dense block than other groups. Groups $5$ and $6$ showed similar language and genre profiles, but their separation suggests the existence of an additional variable governing the probability of edge existence.}
	\label{fig-imdb}
\end{figure}

In order to interpret the output of the biSBM, we downloaded genre and language information from IMDb for each movie. This information, when compared with the partition provided by the model, shows clearly that the existence of an edge is associated with a match between the actor's and the movie's genre and language. Figure~\ref{fig-imdb} shows the bipartite network adjacency matrix $B$, sorted by a degree-corrected partition using $K_{a}=6$, $K_{b} = 6$, and labeled by defining characteristics of each group of movies. Groups $5$ and $6$ are predominantly English movies, while groups $1$, $2$, and $3$ are foreign films, separated by language. Group $4$ on the other hand, is defined not by language, but by genre, consisting of Adult films across many languages.  In the framework of generative models, this correspondence between genre, language, and inferred blocks provides insight into the multiple mechanisms responsible for the existence of edges.

\section{Conclusions}\label{conclusions}
In this paper we have described a stochastic block model for bipartite networks and demonstrated its ability to create and infer bipartite community structure in both degree-corrected and uncorrected regimes. Moreover, we have shown that for bipartite network data, the biSBM is able to find higher likelihood solutions more efficiently than the SBM. Importantly, this bipartite community structure is found without reliance on one-mode projections, and outperforms one-mode projections in all cases tested. 

There are two problems with community detection in one-mode projections, both of which are avoided by the biSBM. First, projections discard information, and second, they create networks composed of overlapping cliques, which often violate the assumptions of the null model underlying the detection method. Using a community detection model that is misspecified for the type of data being analyzed is problematic. The method can fail, or worse, produce a high-scoring partition under the misspecified model. Because methods provide no warnings of either outcome, not only are their results then impossible to correctly interpret, but they may also be misleading, suggesting the presence of strong community structure where there is, in fact, none \cite{good2010}. Whenever possible, the use of one-mode projections should be avoided, with communities instead inferred directly from the original bipartite data.

This point was most evident under our class of synthetic networks which were designed to have ambiguous projections. In these numerical experiments, there existed a community of type $a$ vertices with a high probability of connection to {\it all} type $b$ vertices, and the biSBM substantially outperformed all projection-based methods (Fig.~\ref{fig-synthetic2}B). These results are likely very general, in part because many real-world systems, e.g., a network of documents and the words they contain, contain ubiquitous ``stop'' words that must be removed by hand or heuristically in order for existing methods to work well \cite{lancichinetti2014}. In contrast, the biSBM automatically identifies and classifies such vertices, producing high-quality partitions despite the ubiquitous connectivity of such vertices.

As a brief aside, one-mode projections may be problematic for more than just community detection. For example, it is commonly known that social networks are assortative by degree while most other networks are not, yet the social networks first used to demonstrate this point were all implicitly one-mode projections, such as coauthorship networks \cite{newman2002}. Subsequently, social networks that were not projections were shown to be less assortative or even disassortative \cite{newman2003}. This raises the questions of whether assortativity is due to properties of social networks or due to implicitly projecting from bipartite data, and whether other measures, such as centralities, may also be affected.

The biSBM, in either its degree-corrected or uncorrected form, is mathematically equivalent to a constrained version of the SBM, which allowed for a direct comparison of the two methods. The SBM is a more general model for community detection in networks, but this increased flexibility comes at a cost: when applied to bipartite data, it must learn that these data are bipartite, which causes it to be less efficient at inference, more prone to overfitting, and more likely to produce mixed-type partitions. If the bipartite nature of the network is known ahead of time, this information can and should be utilized. Our results for the biSBM demonstrate that using this information leads to substantially more efficient and more accurate inference.

A subtle point when using the biSBM is the choice of the parameters $K_a$ and $K_b$, which may be chosen independently. This explicit selection of parameters is both an opportunity and a burden, as the increased flexibility allows for modeling imbalanced bipartite networks in which $K_a\not=K_b$, but also requires these parameters to be specified. The choice of these values can be framed as a question of model selection, which compares the likelihoods for different choices while controlling for the added flexibility associated with extra parameters. For SBM-type models, this question is related to, but distinct from the question of choosing the number of communities. (For instance, if $K=K_{a}+K_{b}$, the number of communities in the SBM and biSBM is the same, but the number of free parameters is ${K \choose 2} > K_{a} K_{b}$ for $K>2$.) Techniques for model selection for generative network models like the SBM remain an area of active research. The central difficulty is that the likelihood function's ruggedness makes the standard limiting assumptions inapplicable \cite{yan2013} and  common approaches to comparing models, e.g., AIC and BIC, can produce incorrect decisions. Recent work using likelihood ratio statistics, however, shows promising results~\cite{peel:clauset:2014}, and MDL-based approaches have also been recently developed \cite{shen2011,peixoto2013,peixoto2014}.

The biSBM, and generative models more broadly, fall into a growing set of models in which the generative hypothesis is clear and principled. A strong advantage of such methods is the interpretability of the inferred parameters, as the matrix $\omega$ is informative about hypothetical mechanisms of the underlying processes that generated the data in the first place, e.g., Ref.~\cite{larremore2013}. Mixed-membership stochastic block models \cite{airoldi2008,ball2011}, which assign each vertex a probability distribution over communities, have not yet been formulated for bipartite networks but represent an interesting direction for future work, as do models of edge-weighted networks~\cite{aicher:etal:2013} and non-overlapping edge types~\cite{guimera2012}. Similarly, hierarchical methods~\cite{clauset2008,peixoto2014} could also be adapted to bipartite, $k$-partite, or more complex formulations. Other models have explored structural regularities beyond community structure, where additional model parameters capture inter-group centrality \cite{shen2011}. Given the ubiquity of bipartite and other forms of structured networks, we look forward to the development of more sophisticated generative models the naturally incorporate such auxiliary vertex and edge information.

\section{Acknowledgements}
We thank Leto Peel and Christopher Aicher for helpful conversations.
The project was supported in part by Award Number R21GM100207 (DBL, AC) from the National Institute of General Medical Sciences (NIGMS), and by Grant \#FA9550-12-1-0432 (AC, AZJ) from the U.S.\ Air Force Office of Scientific Research (AFOSR) and the Defense Advanced Research Projects Agency (DARPA). The content is solely the responsibility of the authors and does not necessarily represent the official views of the NIGMS, the National Institutes of Health, AFOSR or DARPA. The funders had no role in study design, data collection and analysis, decision to publish, or preparation of the manuscript. 
An open source and free implementation of these methods is available (see Appendix~\ref{codeanddata}).

\appendix
\renewcommand{\thefigure}{A\arabic{figure}}
\setcounter{figure}{0}
\renewcommand{\thetable}{A\arabic{table}}
\setcounter{table}{0}

\section{Code and Data availability}\label{codeanddata}
Implementations of the biSBM inference code, written by the authors, may be found at \href{http://danlarremore.com/bipartiteSBM}{danlarremore.com/bipartiteSBM}. Southern Women and Malaria data sets are also available at the same web address. IMDb data sets are also available \cite{imdb}.

\section{Southern Women}\label{appendix-southernWomen}
The bipartite SBM described in the text finds the following maximum likelihood partition of the Southern Women network \cite{davis1941}:
{\bf Group A (red)}: Mrs Evelyn Jefferson, 
Miss Laura Mandeville, 
Miss Theresa Anderson, 
Miss Brenda Rogers, 
Miss Charlotte McDowd, 
Miss Frances Anderson, 
Miss Eleanor Nye, 
Miss Pead Oglethorpe, 
Miss Ruth DeSand.
{\bf Group B (blue)}: Miss Verne Sanderson, 
Miss Myra Liddell, 
Miss Katherine Rogers, 
Mrs Sylvia Avondale, 
Mrs Nora Fayette, 
Mrs Helen Lloyd, 
Mrs Dorothy Muchison, 
Mrs Olivia Carleton, 
Mrs Flora Price.
{\bf Group X (orange)}: Jun10, Jan23, Apr07, Nov21, Aug03.
{\bf Group Y (purple)}: Mar15, Sep16, Apr08.
{\bf Group Z (green)}:  Jun27, Mar02, Apr12, Sep25, Feb25, May19.

\begin{figure}[h]
	\centering
	\includegraphics[width=1.0\linewidth]{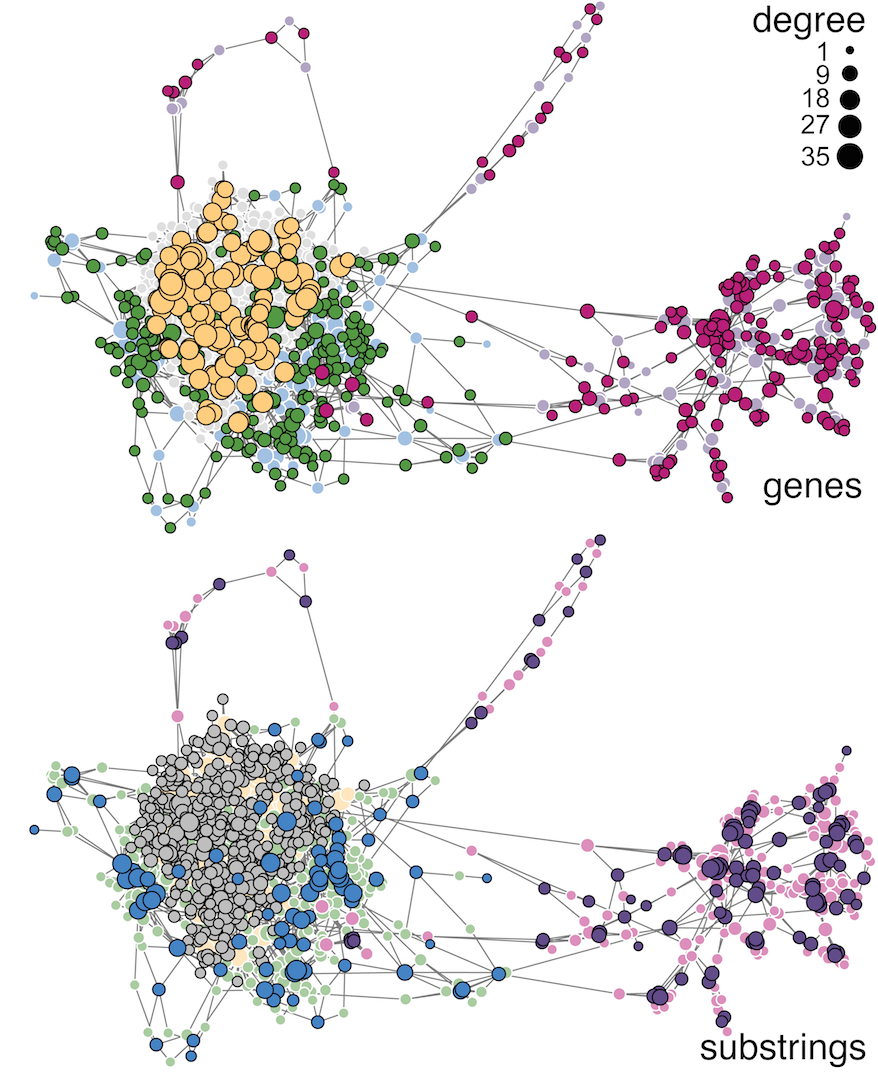}
	\caption{(Color online) Without degree correction, the biSBM tends to find groups that have a similar degree, leading to unexpected and unintuitive partitions of networks with broad or heterogenous degree distributions (as in \cite{karrer2011}). The maximum likelihood partition without degree correction is shown above for the Malaria network, with vertex sizes corresponding to degree. The networks plotted in both panels are identical except for the type of vertices highlighted. The degree-corrected partition is shown in Fig.~\ref{fig-malaria}.}
	\label{fig-malariaNODC}
\end{figure}


\begin{thebibliography}{99}
\bibitem{note-types} Given a bipartite network consisting of a single component, one can arbitrarily determine vertex types. This is possible but appropriate only if $K_a = K_b$. For networks of multiple components and unknown types, modified likelihood maximization approaches are conceivable, but we offer none here.
\bibitem{note-k1relaxation} An exception to this is made when $K_a$ or $K_b$ is equal to one, in which case some vertices will have no possible moves and are accordingly skipped.
\bibitem{noteOnDegrees} Not unlike other generative network models, there are restrictions on allowable parameters. In this case, we fix $\omega$ and let $\theta$ vary by some multiplicative constant for each community, so that we may plant heterogeneous degrees in $\theta$ without over- or mis-specifying $\omega$.
\end{thebibliography}
\end{document}